\documentclass[journal=jpclcd,manuscript=article]{achemso}
\setkeys{acs}{articletitle=true}

\usepackage[version=3]{mhchem} 

\usepackage{amsmath}
\usepackage{amsfonts}
\usepackage{amssymb}
\usepackage{graphicx}
\usepackage{dcolumn}
\usepackage{bm}
\usepackage{subcaption}
\usepackage{algorithm} 
\usepackage{algpseudocode} 
\usepackage{xr}
\usepackage{hyperref}
\usepackage{braket}
\usepackage{xcolor}

\makeatletter
\newcommand*{\addFileDependency}[1]{
  \typeout{(#1)}
  \@addtofilelist{#1}
  \IfFileExists{#1}{}{\typeout{No file #1.}}
}
\makeatother

\usepackage[capitalize]{cleveref}
\crefname{figure}{Fig.}{Figs.}
\Crefname{figure}{Figure}{Figures}
\crefname{table}{Tab.}{Tabs.}
\Crefname{table}{Table}{Tables}
\crefname{equation}{Eq.}{Eqs.}
\Crefname{equation}{Equation}{Equations}
\crefname{section}{Sec.}{Secs.}
\Crefname{section}{Section}{Sections}
\usepackage{xcolor}

\usepackage{array}
\newcommand{\PreserveBackslash}[1]{\let\temp=\\#1\let\\=\temp}
\newcolumntype{C}[1]{>{\PreserveBackslash\centering}p{#1}}
\newcolumntype{R}[1]{>{\PreserveBackslash\raggedleft}p{#1}}
\newcolumntype{L}[1]{>{\PreserveBackslash\raggedright}p{#1}}

\usepackage{multirow}

\author{Fabijan Pavo\v{s}evi\'{c}}
\email{fpavosevic@gmail.com}
\affiliation{Center for Computational Quantum Physics, Flatiron Institute, 162 5th Ave., New York, 10010  NY,  USA}

\author{Robert L. Smith}
\affiliation{Center for Computational Quantum Physics, Flatiron Institute, 162 5th Ave., New York, 10010  NY,  USA}
\alsoaffiliation{ Department of Chemistry, Virginia Tech, Blacksburg, Virginia 24061, U.S.A. }

\author{Angel Rubio}
\email{angel.rubio@mpsd.mpg.de}
\affiliation{Center for Computational Quantum Physics, Flatiron Institute, 162 5th Ave., New York, 10010  NY,  USA}
\altaffiliation{Max Planck Institute for the Structure and Dynamics of Matter and
Center for Free-Electron Laser Science \& Department of Physics,
Luruper Chaussee 149, 22761 Hamburg, Germany}

\title[]
  {Cavity Click Chemistry: Cavity-Catalyzed Azide–Alkyne Cycloaddition}

\begin{document}



\begin{tocentry}
\begin{figure}[H]
	\begin{center}
		\includegraphics[width=1.7in]{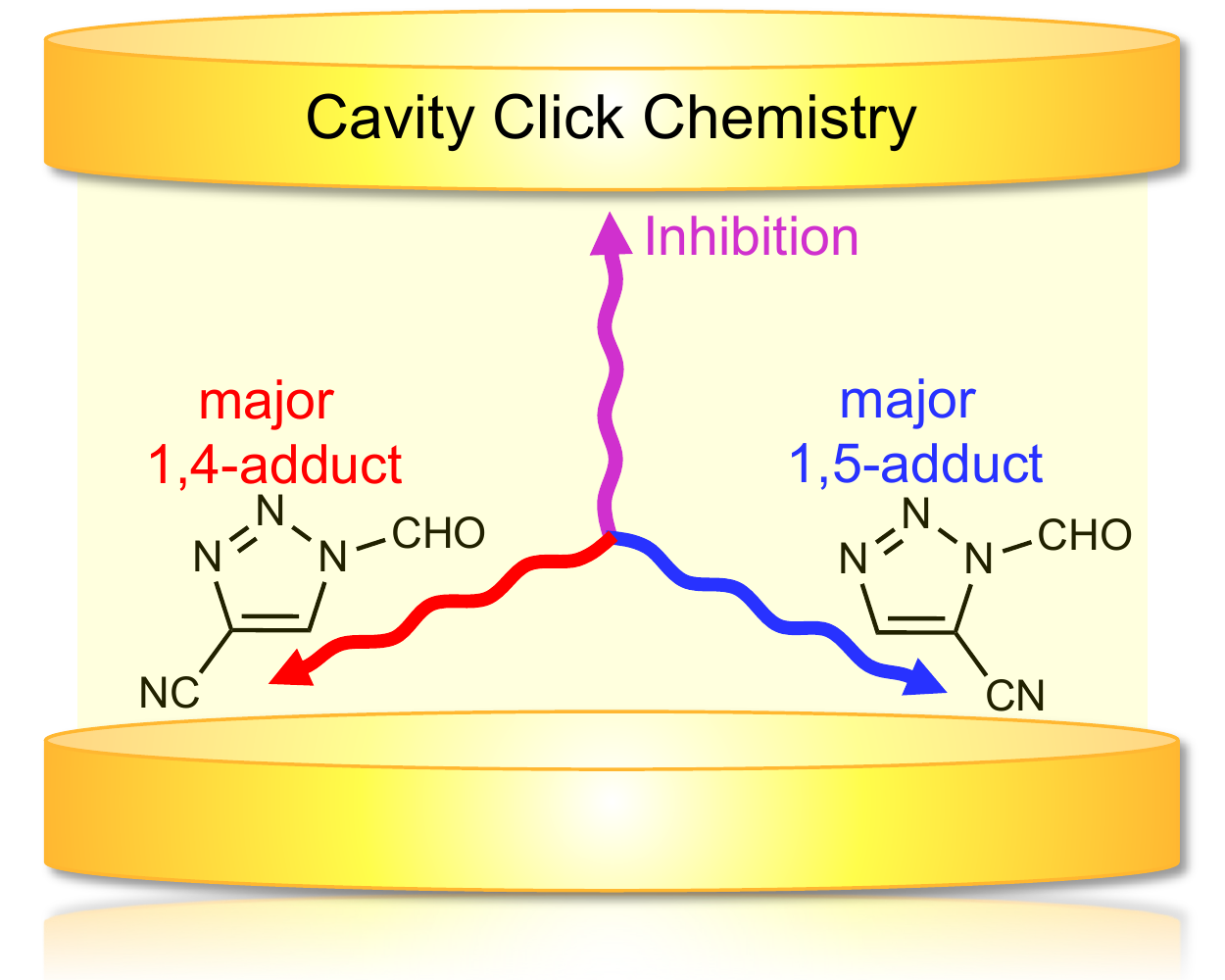}

	\end{center}
\end{figure}
\end{tocentry}

\begin{abstract}
Click chemistry, which refers to chemical reactions that are fast, selective, and with high product yields, has become a powerful approach in organic synthesis and chemical biology. Due to the cytotoxicity of the transition metals employed in click chemistry reactions, a search for novel metal-free alternatives continues. Herein we demonstrate that an optical cavity can be utilized as a metal-free alternative in click chemistry cycloaddition reaction between cyanoacetylene and formylazide using the quantum electrodynamics coupled cluster (QED-CC) method. We show that by changing the molecular orientation with respect to the polarization of the cavity mode(s), the reaction can be selectively catalyzed to form a major 1,4-disubstituted or 1,5-disubstituted product. This work highlights that a cavity has the same effect on the investigated cycloaddition as the transition metal catalysts traditionally employed in click chemistry reactions. We expect our findings to further stimulate research in cavity-assisted click chemistry reactions.
\end{abstract}

\maketitle


Click chemistry refers to a group of highly reactive, selective, high-yield, and bioorthogonal chemical reactions that occur under mild reaction conditions.~\cite{kolb2001click} Due to this set of stringent criteria that defines an ``ideal chemical reaction", it has emerged in the last years as one of the most powerful tools in drug discovery,~\cite{thirumurugan2013click}
chemical biology,~\cite{fantoni2021hitchhiker} and materials science.~\cite{taiariol2021click} Although several different reactions have been identified to fit the concept of click chemistry, the 1,3-dipolar cycloaddition is the most utilized example.~\cite{kolb2001click,breugst2020huisgen} In particular, the synonym for the click chemistry is the copper-catalyzed 1,3-dipolar Huisgen cycloaddition reaction of azides with alkynes (CuAAC) in the synthesis of chemically and biologically important triazoles.~\cite{tornoe2002peptidotriazoles,rostovtsev2002stepwise} In presence of the copper core, the 1,3-dipolar cycloaddition reactions at room temperature and in various solvents are accelerated by $\sim$10$^7$-fold while forming only 1,4-disubstituted triazoles,~\cite{rostovtsev2002stepwise,himo2005copper} whereas in the corresponding thermal reactions both 1,4-disubstituted and 1,5-disubstituted triazoles are formed at a slow rate.~\cite{rostovtsev2002stepwise,breugst2020huisgen} Analogous to the CuAAC, the ruthenium-catalyzed azide-alkyne cycloaddition (RuAAC) promotes the formation of 1,5-disubstituted triazoles.~\cite{zhang2005ruthenium} However, due to the cytotoxicity of the employed transition metal catalysts that control the reaction selectivity, these approaches have limited applicability in biological settings. This has prompted the development of alternative metal-free click chemistry strategies.~\cite{sletten2011mechanism}

Another not yet explored option for catalyzing or imparting selective control over click chemical reactions can be achieved through polaritonic chemistry,~\cite{sidler2020polaritonic,schafer2022shining,sun2022suppression,mandal2022theoretical} in which the strong coupling between quantized light and molecules can be employed to modulate chemical reactions.~\cite{ebbesen2016hybrid,herrera2016cavity,ribeiro2018polariton,ruggenthaler2018quantum,garcia2021manipulating,hubener2021engineering,sidler2022perspective,ruggenthaler2022understanding} Several groups of chemical reactions under the strong light-matter coupling regime have been studied so far, such as nucleophilic substitutions,~\cite{thomas2016ground,thomas2019tilting} isomerization,~\cite{sun2022suppression} or electrocyclization,~\cite{sau2021modifying} however, none of them is within the realm of click chemistry. These strong light-matter interactions are usually created inside optical cavities, where photonic quantum vacuum fluctuations couples with a molecular system giving rise to the molecular polaritons. Because the polaritonic properties can be modulated on demand by changing the field inside an optical cavity, this approach offers a non-invasive way to catalyze,~\cite{lather2019cavity} inhibit,~\cite{thomas2016ground} or to control reactions selectivity.~\cite{thomas2019tilting} To understand and enhance the experimental design in polaritonic chemistry, the development of accurate and efficient {\it ab initio} methods is required. Recently developed quantum electrodynamics coupled cluster (QED-CC) methods,~\cite{haugland2020coupled,deprince2021cavity,Pavosevic2021} where both electrons and photons are treated quantum mechanically on equal footing, offers a robust and systematically improvable way for studying cavity mediated chemical reactions.~\cite{haugland2020coupled,deprince2021cavity,Pavosevic2021,pavosevic2021cavity,pavovsevic2022wavefunction,pavovsevic2022catalysis} 

Herein we show that the strong light-matter interactions created in an optical cavity offer a metal-free approach to modulating reaction rates and product selectivity of click chemistry reactions. To that end, we employ the QED-CC method~\cite{haugland2020coupled,deprince2021cavity,Pavosevic2021} for obtaining an insight into the effect of strong light-matter interaction on catalysis, inhibition, and selective control in the prototypical 1,3-dipolar Huisgen cycloaddition reaction between cyanoacetylene and formylazide. As we will show in the remainder of this work, the corresponding reaction is significantly inhibited if the cavity mode is polarized along the bond-forming direction. In contrast, the polarization of the cavity modes in the remaining two directions catalyzes the reaction to give major 1,4-disubstituted or 1,5-disubstituted triazoles.

The light-matter interaction between a molecular system and a quantized light can be described by the Pauli-Fierz Hamiltonian~\cite{ruggenthaler2018quantum}, which under the dipole approximation, in the coherent state basis~\cite{haugland2020coupled}, and for a single-mode cavity reads as (note that extension to multiple modes is straightforward~\cite{pavovsevic2022catalysis}):
\begin{equation}
\begin{aligned}
    \label{eqn:PF_Hamiltonian_CSB}
    \hat{H}=\hat{H}^{\text{e}}+\omega b^{\dagger}b-\sqrt{\frac{\omega}{2}}(\boldsymbol{\lambda} \cdot \Delta\boldsymbol{d})(b^{\dagger}+b)+\frac{1}{2}(\boldsymbol{\lambda} \cdot \Delta\boldsymbol{d})^2
\end{aligned}
\end{equation}
Here, $\hat{H}^{\text{e}}$ corresponds to the electronic Hamiltonian within the Born-Oppenheimer approximation (description beyond Born-Oppenheimer approximation can also be incorporated~\cite{Hammes-Schiffer19_338,Hammes-Schiffer20_4222,pavosevic2021multicomponent}). The second term represents the Hamiltonian for the cavity mode with fundamental cavity frequency $\omega$ where $b^{\dagger}$ and $b$ are bosonic creation and annihilation operators, respectively. The last two terms of Eq.~\ref{eqn:PF_Hamiltonian_CSB} are the dipolar coupling (accounting for interactions between electronic and photonic degrees of freedom) and the dipole self-energy (accounting for the quantum vacuum fluctuations), respectively. Within these two terms, $\boldsymbol{\lambda}$ and $\Delta\boldsymbol{d}=\boldsymbol{d}-\langle\boldsymbol{d}\rangle$ are the strong light-matter coupling strength vector and the molecular dipole moment operator shifted by its mean-field expectation value, respectively. The quantum electrodynamics coupled cluster (QED-CC) interaction energy, $E_{\text{QED-CC}}$, of a molecule embedded inside a cavity is obtained by solving the Schr\"odinger equation
\begin{equation}
\begin{aligned}
    \label{eqn:Schrodinger_eqn}
    \hat{H}|\Psi_{\text{QED-CC}}\rangle=E_{\text{QED-CC}}|\Psi_{\text{QED-CC}}\rangle
\end{aligned}
\end{equation}
In this equation, $|\Psi_{\text{QED-CC}}\rangle=e^{\hat{T}}|\Psi_{\text{QED-HF}}\rangle$ is the QED-CC ground state wave function~\cite{haugland2020coupled} where $\hat{T}$ is the cluster operator that incorporates correlated interactions between quantum particles (electrons and photons), and $|\Psi_{\text{QED-HF}}\rangle$ is the reference QED-Hartree-Fock wave function.~\cite{haugland2020coupled} In this work, we use the QED-CC method in which the cluster operator includes up to single and double electronic transitions and the creation of a single photon, along with interactions of up to two electrons with a single photon.~\cite{haugland2020coupled} In the remainder of this work, we will refer to this method as QED-CCSD.

\begin{figure*}[ht!]
  \centering
  \includegraphics[width=6.5in]{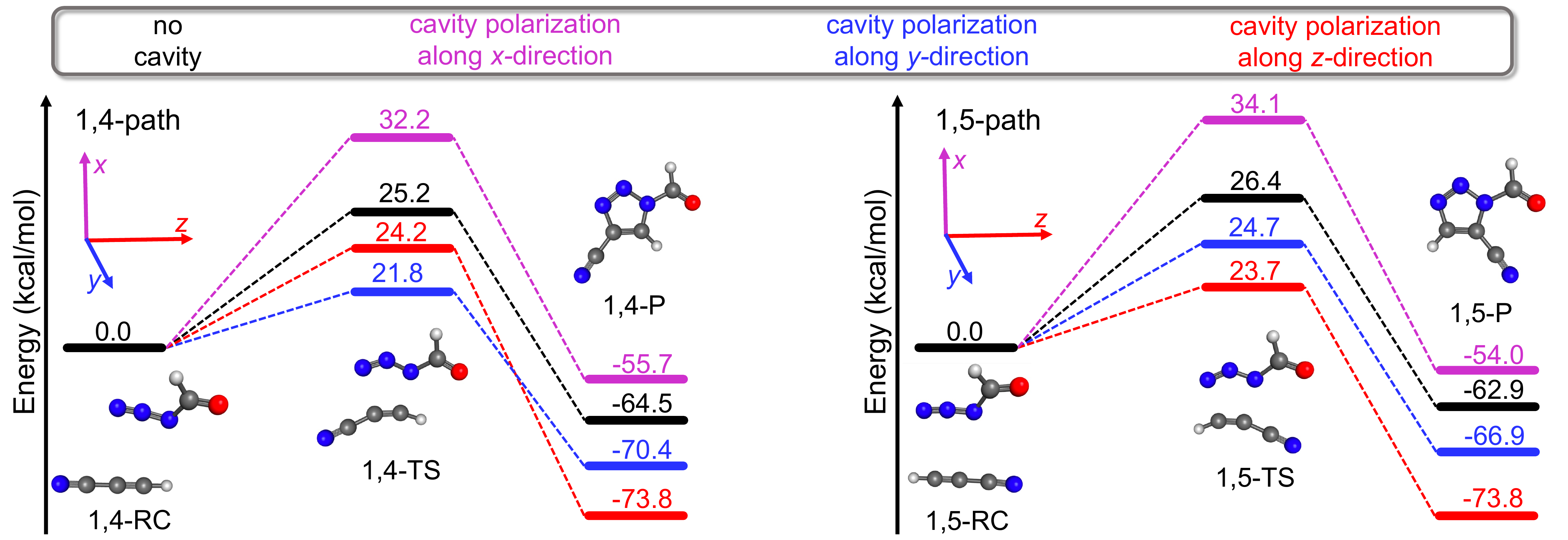}
  \caption{Reaction diagram of 1,3-dipolar Huisgen cycloaddition reaction between cyanoacetylene and formylazide for 1,4-path (left panel) and 1,5-path (right panel) calculated with the QED-CCSD/cc-pVDZ method. The QED-CCSD calculations employ the cavity parameters $\omega=1.5$~eV and $|\lambda |=0.1$~a.u. with the photon mode polarized in the $x$ (magenta, along the forming C-N bond), $y$ (blue, perpendicular to the triazole ring plane), and $z$ (red, in the plane of the triazole ring) molecular directions. The no cavity (CCSD) calculations are in black. The panels contain the structures of the reaction complex (RC), transition state (TS), and product (P), along with the molecular coordinate frame.}
  \label{fig:energy_diagram}
\end{figure*}

All the QED-CCSD calculations described throughout this work employ the cc-pVDZ basis set~\cite{dunning1989gaussian} and were performed on the geometries optimized at the CCSD/cc-pVDZ level of theory (provided in the Supporting Information) using the Q-Chem quantum chemistry software.~\cite{epifanovsky2021software} We note that the effect of the cavity on the structural relaxations is not considered here, and we assumed that molecular geometries do not dramatically change due to strong light-matter coupling, as shown previously.~\cite{pavovsevic2022catalysis} 
The source code for the QED-CCSD method is available in Ref.~\citenum{pavosevic2021cavity}. The orientation of the structures is kept constant along the reaction path such that the forming bond is oriented along $x$-direction and in the plane of the forming triazole ring (see Fig.~\ref{fig:energy_diagram}). 

\begin{table}
\caption{Activation energy barrier ($E_\text{a}$)\textsuperscript{a} and reaction energy ($\Delta E$)\textsuperscript{b} (in kcal/mol) for 1,3-dipolar Huisgen cycloaddition reaction between cyanoacetylene and formylazide along the 1,4-path and 1,5-path calculated with the QED-CCSD/cc-pVDZ method with cavity frequency of 1.5~eV and light-matter coupling of 0.1~a.u. Values in parenthesis correspond to calculations for light-matter coupling of 0.2~a.u.}
\begin{tabular}{c|c|C{2cm}|c|c}
\hline
&  path  & $E_\text{a}$ & ratio\textsuperscript{c}& $\Delta E$\\\hline\hline              
\multirow{2}{*}{no cavity\textsuperscript{d}} & 1,4-path & 25.2 & \multirow{2}{*}{88:12} & -64.5\\
                 & 1,5-path & 26.4 &  & -62.9\\\hline
\multirow{2}{*}{$x$-direction} & 1,4-path & 32.2 (40.1) & \multirow{2}{*}{96:4 (99:1)} & -55.7 (-47.4)\\
                  & 1,5-path & 34.1 (42.9) & & -54.0 (-45.5)\\\hline
\multirow{2}{*}{$y$-direction} & 1,4-path & 21.8 (15.5) & \multirow{2}{*}{99:1 (100:0)} & -70.4 (-83.4)\\
                  & 1,5-path & 24.7 (21.1) & & -66.9 (-76.3)\\\hline
\multirow{2}{*}{$z$-direction} & 1,4-path & 24.2 (21.9) & \multirow{2}{*}{30:70 (0:100)} & -73.8 (-87.6)\\
                  & 1,5-path & 23.7 (18.4) & & -73.8 (-90.6) \\\hline\hline      
\end{tabular}
\\
\raggedright \textsuperscript{a}\small Calculated as the energy difference between the transition state and the reaction complex.\\

\textsuperscript{b}\small Calculated as the energy difference between the product and the reaction complex.\\

\textsuperscript{c}\small 1,4-product/1,5-product ratio for kinetically controlled reaction.\\


\textsuperscript{d}\small No cavity indicates the conventional electronic structure CCSD/cc-pVDZ result.\\

\label{table:table}
\end{table}

Figure~\ref{fig:energy_diagram} shows the reaction energy diagram for the 1,3-dipolar Huisgen cycloaddition reaction between cyanoacetylene and formylazide along the 1,4-path (left panel) and 1,5-path (right panel). The activation energies (difference between the transition state and the reaction complex energies) and the reaction energies (difference between the product and the reaction complex energies) are calculated with the conventional electronic CCSD/cc-pVDZ method (black) which we also refer to as `no cavity', and with the QED-CCSD/cc-pVDZ method with the cavity mode polarized in the molecular $x$ (magenta), $y$ (blue), and $z$ (red) directions with respect to the molecular coordinate frame indicated in Fig.~\ref{fig:energy_diagram}. In the following, we investigate the effect of the optical cavity on the reactions selectivity under the kinetic control as well as the thermodynamic stabilities of products which are essential principles of the click chemistry reactions. In particular, we focus on changes in the relative product ratio along the 1,4-path and 1,5-path in the case of kinetically controlled reactions and on changes in reaction energy due to a cavity. The product ratio for a kinetically controlled reaction is calculated by plugging in the difference between activation energies along two different reaction paths into the Arrhenius equation~\cite{arrhenius1889dissociationswarme} at room temperature ($T=298.15$~K). The calculated product ratios for 1,4-path and 1,5-path in case of kinetically controlled reaction are given in Table~\ref{table:table}. 

As given in Fig.~\ref{fig:energy_diagram} (in black) and Table~\ref{table:table} (no cavity), the activation energy for the gas phase reaction calculated with the CCSD method along the 1,4-path is lower by 1.2~kcal/mol than for the 1,5-path. Due to this difference, the 1,4-product:1,5-product ratio for the kinetically controlled reaction is 88:12. We note that for the CCSD(T) method, in which the triple electronic excitations are treated perturbatively, this difference is 0.8~kcal/mol and the 1,4-product:1,5-product ratio is 79:21. This is in good agreement with an accurate theoretical estimate reported in Ref.~\citenum{jones2008predictions}, that determines this activation energy difference at 0.9~kcal/mol along with the 1,4-product:1,5-product ratio of 82:18. Therefore, in this thermal reaction under the kinetic control, both 1,4-disubstituted and 1,5-disubstituted products will be formed. Moreover, the reaction without a catalyst will proceed slowly due to a high activation energy along both paths.~\cite{liang2011copper} 

Next, we consider the same thermal reaction embedded inside an optical cavity with photon mode frequency $\omega=1.5$~eV and light-matter coupling constant $|\lambda|=0.1$~a.u. Both are within the range of current experimental setups.~\cite{Eizner2019,wu2021} We note that the latter can be enhanced by exploiting collectivity effect~\cite{garcia2021manipulating}, ensemble of nearby emitters~\cite{schutz2020ensemble}, multimode cavities~\cite{vaidya2018tunable}, or utilizing the strong light-confinement in picocavities.~\cite{baumberg2022picocavities} In the case where the cavity mode is polarized in the molecular $x$-direction, which corresponds to the C-N bond forming direction and in the plane of the forming triazole ring (see Fig~\ref{fig:energy_diagram}), the reaction activation energy increases by 7.0~kcal/mol and 7.7~kcal/mol along the 1,4-path and 1,5-path, respectively. Due to such a high increase in the reaction activation energy, the reaction's rate along both paths decreases by $\sim$10$^5$ orders at room temperature. As a result, the barrier along the 1,5-path is increased more than that of the 1,4-path, resulting in the calculated 1,4-product:1,5-product ratio of 96:4. 

Now if the cavity mode is polarized along the molecular $y$-direction, which is perpendicular to the plane of the forming triazole ring, the activation energy along the 1,4-path is lowered by 3.4~kcal/mol, whereas for the 1,5-path by 1.8~kcal/mol. Such a decrease in the activation energy along the 1,4-path corresponds to an increase in the reaction rate by 310 times. Additionally, due to this uneven change in the activation energy, the 1,4-product:1,5-product ratio is 99:1. Additionally, we have performed the same calculation with the light-matter coupling strength of $|\lambda|=0.2$~a.u. The results are provided in parenthesis of Table~\ref{table:table}. For this value of the light-matter coupling strength with light polarized along the molecular $y$-direction, the reaction activation energy decreases by 9.7~kcal/mol along the 1,4-path, increasing the reaction rate by $\sim$10$^7$ orders as well as in 100:0 1,4-product:1,5-product ratio. This indicates that a cavity with the mode polarized along molecular $y$-direction has the same effect as adding the copper catalyst,~\cite{himo2005copper} i.e., increasing the reaction rate by $\sim$10$^7$ orders and yielding only 1,4-disubstituted product.~\cite{himo2005copper}

Lastly, when the cavity mode is polarized along the molecular $z$-direction, the activation energy along the 1,4-path is decreased by 1.0~kcal/mol, whereas along the 1,5-path, the activation energy is decreased by 2.7~kcal/mol. This leads to the inversion of the reaction barriers such that the activation energy along the 1,5-path is lower by 0.5~kcal/mol than the activation energy along the 1,4-path. Due to this, the calculated 1,4-product:1,5-product ratio is 30:70. Moreover, the reaction rate along the 1,5-path increases by 96 times. For the light-matter coupling strength of $|\lambda|=0.2$~a.u., the activation energies are decreased by 3.3~kcal/mol and 8.0~kcal/mol along the 1,4-path and 1,5-path, respectively, giving the 1,4-product:1,5-product ratio of 0:100. Therefore, the polarization of a cavity along the molecular $z$-direction in a strong light-matter coupling configuration has a same effect on the reaction as the ruthenium catalyst.~\cite{zhang2005ruthenium}

Another important feature of the click chemistry reactions is a high reaction energy, i.e., the thermodynamic stability of the products. Next, we discuss the effect of a cavity on the reaction energy. For the no cavity reaction (CCSD), a large negative reaction energy (see Fig.~\ref{fig:energy_diagram} and Table~\ref{table:table}) indicate that the reaction along both 1,4-path and 1,5-path is highly exothermic.~\cite{jones2008predictions} The same reaction along 1,4-path and 1,5-path inside an optical cavity with the mode polarized along the molecular $x$-direction becomes less exothermic by 8.8~kcal/mol and 8.9~kcal/mol, respectively. If the cavity mode is now polarized along the molecular $y$-direction, the reaction along both 1,4-path and 1,5-path becomes more exothermic by 5.9~kcal/mol and 4.0~kcal/mol, respectively. For a stronger light-matter coupling strength of $|\lambda|=0.2$~a.u., the reaction along both 1,4-path and 1,5-path becomes more exothermic by 18.9~kcal/mol and 13.4~kcal/mol, respectively. Finally, the most significant effect of the cavity mode(s) on the reaction energy is observed when the cavity mode is polarized along the molecular $z$-direction. Under these conditions, the reaction along the 1,4-path and 1,5-path is stabilized by 9.3~kcal/mol and 10.9~kcal/mol, respectively. The strong light-matter coupling strength of $|\lambda|=0.2$~a.u. makes the reaction even more exothermic by 23.1~kcal/mol and 27.7~kcal/mol along both paths, respectively. Therefore, this indicates that a cavity can be utilized to change the reaction products' thermal stability and modulate the reaction's reversibility.

Click chemistry is grounded on two important principles, the kinetic control of selectivity and the high thermodynamic stability of products. In this work, we have theoretically demonstrated that an optical cavity can be utilized for modulating the product's selectivity under kinetic control and reactions reversibility of 1,3-dipolar cycloaddition click chemistry reaction between cyanoacetylene and formylazide. In particular, we have shown that a cavity with the light mode polarized in the C-N bond-forming direction inhibits the overall reaction. Conversely, when the cavity mode is polarized in the remaining two molecular directions denoted as the Cartesian $y$ and $z$, the reaction is catalyzed along both 1,4-path and 1,5-path, respectively, leading to selective formation of major 1,4-disubstituted or 1,5-disubstituted products. These findings indicate that a cavity with the light polarized along these two directions offers a metal-free alternative to the copper or ruthenium catalysts~\cite{himo2005copper,zhang2005ruthenium} with the same rate increasing and stereoselective control effects on the reaction in a strong, but achievable, light-matter coupling regime. Lastly, we have shown that for a cavity with $y$ and $z$ oriented fields, the reaction along both reaction paths becomes significantly more exothermic relative to no cavity case. Due to the great importance of click chemistry in drug discovery, chemical biology, and proteomic applications~\cite{kolb2001click}, we expect that our findings will further stimulate both experimental and theoretical research in cavity enhancement of click chemistry reactions. 

\begin{acknowledgement}
We acknowledge financial support from the Cluster of Excellence 'CUI: Advanced Imaging of Matter'- EXC 2056 - project ID 390715994 and SFB-925 "Light induced dynamics and control of correlated quantum systems" – project 170620586  of the Deutsche Forschungsgemeinschaft (DFG) and Grupos Consolidados (IT1453-22). We also acknowledge support from the Max Planck–New York Center for Non-Equilibrium Quantum Phenomena. The Flatiron Institute is a division of the Simons Foundation.
\end{acknowledgement}

\noindent\textbf{Supporting Information Available}: The supporting information includes: Cartesian coordinates of the optimized geometries.

\noindent\textbf{Conflict of interest}\\
The authors declare no conflict of interest.

\linespread{1}\selectfont
\bibliography{references}{}

\end{document}